# Self-Diffusion of Iron in $L1_0$ FePd films Upon He-irradiation (alternative title: An alternative for diffusion studies in isotopic FePd multilayers)

D. G. Merkel, F. Tanczikó, Sz. Sajti, M. Major, Cs. Fetzer and L. Bottyán
KFKI Research Institute for Particle and Nuclear Physics, P.O.B 49, H 1525, Budapest, Hungary

A. Kovács
MTA Research Institute for Technical Physics and Materials Science (MFA), P.O.B 49, H 1525, Budapest, Hungary

R. Rüffer, S. Stankov
European Synchrotron Radiation Facility, BP 220, 38043 Grenoble CEDEX 9, France

**ABSTRACT**

The continuous need towards improving the capacity of magnetic storage devices requires materials with strong perpendicular magnetic anisotropy. FePd, CoPd and their Co(Fe)Pt counterparts very attractive candidate for such purposes. The magnetic properties of these films are largely dependent on the orientation and local distribution of the $L1_0$ FePd phase fraction which are mainly controlled by diffusion processes and involve diffusion paths of a few angstroms. Highly ordered as well as disordered epitaxial isotope-periodic $^{57}$FePd/$^{nat}$FePd (001) thin films were prepared by molecular-beam epitaxy on MgO(001) substrate. Short range diffusion of different phases in FePd thin film induced by room temperature 130 keV $He^+$ irradiation was investigated at fluences up to $30\times10^{15}$ ions/cm$^2$. Conversion electron Mössbauer spectroscopy and synchrotron Mössbauer reflectivity experiments showed that the inter-atomic diffusion across $^{57}$FePd/$^{nat}$FePd interface occur mainly via the iron rich regions. The ratio of the diffusion length in the $L1_0$, fcc and iron rich structure are 1:1.4:5.6 respectively. Assuming that the diffusion coefficient in the *fcc* phase is between the diffusion coefficient of the $L1_0$ phase in the crystallographic *c* direction and perpendicular to it, the diffusion coefficient in the *c*-direction of the $L1_0$ phase is found more than 1.9 times lower than in the *a*-*b* plane.

## I INTRODUCTION

Due to their high perpendicular magnetic anisotropy, (PMA), $L1_0$ (CuAu(I)-type) ordered FePd as well as FePt and CoPt are candidate materials for future ultra-high density magnetic recording [1,2,3]. PMA has been reported to originate in these materials from growth-induced heterogeneity, compressive strain, short-range order driven segregation or magnetically induced phase separation, but was not found to be linked to the appearance of any long range ordered phase. At room temperature, $Fe_{1-x}Pd_x$ exhibits equilibrium $L1_0$ structure in the composition range $0.5 < x < 0.6$, with alternating Fe and Pd planes along the [001] direction. In the $L1_0$ FePd structure the short and long lattice parameter ratio (former along the [001] axis) varies between 0.96 and 0.97. The anisotropy energy is in range of $10^7$erg/cm$^3$ [4,5,6] which is of primary interest for magnetic storage applications since it allows nano-size magnetic volumes to remain ordered against thermal agitation. The ordered $L1_0$ to disordered fcc phase transition is driven by short range diffusion.

Atomic diffusion governs the structural relaxation and associated changes in the physical and magnetic properties in these alloys. Depth profiling techniques like radioactive tracer and secondary ion mass spectrometry are most widely used for such studies. However, the depth resolution of these techniques is often limited to a few houndred angströms. This is a severe limitation, since in metastable systems such large diffusion lengths may be difficult to achieve without fundamental structural transformations. Dynamic techniques like Mössbauer spectroscopy or nuclear forward scattering of synchrotron radiation, on the other hand, are limited by the data acquisition time and the time window over which they are sensitive to atomic diffusion. For example, diffusivities accessible using $^{57}$Fe Mössbauer spectroscopy are in the range of $10^{-12}$ to $10^{-13}$ m$^2$/s [7] which is not enough to follow diffusion in atomic range.

In studies of chemically modulated multilayers x-ray diffraction (XRD) and x-ray reflection (XRR) methods are commonly used. By XRR, it is possible to study diffusion lengths below the detection limit achievable by the sectioning and profiling techniques [8,9,10,11,12]. Despite the difficulties that arise due to the concentration dependence of the diffusion coefficient, these techniques are widely used to study interdiffusion in compositionally heterogeneous multilayers, it is not possible to follow self-diffusion by these techniques in chemically homogenous materials since the solely electronic interaction with the x-ray photons produces no contrast between the adjacent layers.

To investigate self diffusion, an isotope sensitive technique is required. Conventional x-ray techniques don't have this feature but neutrons may have significant difference in scattering cross section for the different isotopes of the same element. Neutron reflectivity is a suitable non-destructive method to study self-diffusion in a chemically homogenous isotope multilayer with diffusion lengths of the order of a few angstrøms [13,14]. A recently emerged similar isotope sensitive non destructive method for studying self diffusion is Synchrotron Mössbauer Reflectivity (SMR) [15]. Due to the nuclear (hyperfine) interaction between the atomic nuclei and the highly monochromatized synchrotron radiation (tuned to the transition energy of the resonant isotope, hereafter $^{57}$Fe), Mössbauer isotopes have very large scattering amplitude as compared to that of non-Mössbauer-isotopes of the same material providing an isotope contrast between the adjacent layers of the sample. This results in Bragg peaks in the reflectivity pattern, related to the period of the periodic multilayer sample. The sensitivity of these methods are suitable for the layer thickness used in our case. From the variation of the Bragg peak intensity, valuable information can be gathered on the change of the isotope depth profile, and consequently on the interlayer diffusion. However, isotope sensitive neutron and Mössbauer reflectometry has been already used by different groups [16,17] they assumed that the sample structure remained unchanged during the heat treatment thus neglecting the effect of phase evolution.

Low energy(~130keV) $He^+$ implantation features low collision cross section with the intention to avoid defect interactions and small energy transfer to minimize recoil displacement. However, the beam energy is sufficient for the ions to pass through the layer and stop deep (~1.5 μm) in the substrate, leaving a rather homogeneous defect distribution in the (65 nm thick) film to relax. According to SRIM [18] simulations each incoming ion generates ~6 displacements on average, which mainly generate vacancies and equal number of interstitials According to these simulations, the probability of sputtering by these $He^+$ ions is negligible ($< 10^{-5}$). This kind of low energy ion implantation doesn't induce collision cascades in the layer and results similar effect as heat treatment.

In this paper we use SMR method to study the self-diffusion of Fe in an isotope periodic natFePd/$^{57}$FePd multilayer film following room temperature low energy $He^+$ irradiation of various doses. Isotope layers with originally sharp interfaces were used to avoid chemical effects of diffusion. X-ray diffraction (XRD), conversion electron Mössbauer spectroscopy (CEMS), high resolution transmission electron microscopy (HRTEM), completed by selected area electron diffraction (SAED) and SMR were used to characterize the samples and to follow the irradiation induced processes in the samples.

## II THEORETICAL CONSIDERATION

SMR experiments can give information on the atomic movement below the nanometer scale, which is indispensable to understand self-diffusion-controlled processes. The $^{57}$Fe concentration in one isotope layer of $^{nat}Fe_{47}Pd_{53}$/ $^{57}Fe_{47}Pd_{53}$ system can be expressed in a Fourier series [11,19]

$$C(z) = \sum_n C_n \exp(ik_n z) \tag{1}$$

Where $k_n=2n\pi/\Lambda$ with $\Lambda$ being the periodicity of the multilayer and $z$ the depth of the sample. When atomic diffusion takes place in the sample by annealing or irradiation the magnitude of the Fourier component $C_n$ decays with time. If the atomic migration by annealing results diffusion in the sample then we can use the 1D Fick equation. Substituting (1) into this equation one obtain the following equation.

$$C_n = C_{0n} \exp\left[-k_n^2 Dt\right] \tag{2}$$

Where $D$ and $t$ are the diffusivity at a given temperature, and the retention time, respectively. If the atomic migration is diffusion then $D$ depends from temperature Arrhenius like. If the atomic migration is irradiation induced then $D$ depends from the energy loss of the ion beam for a given depth.

T. Mizoguchi et al.[20] derived a formula, which relates the intensity of the *n*-th order Bragg peak in the reflectivity pattern at $t$ = 0 ($I_0$) to that at time $t$, according to the following expression:

$$\log\left(\frac{I(t)}{I_0}\right) = -\frac{n^2\pi^2}{\Lambda^2} D(T)t \, , \tag{3}$$

where log stands for the natural logarithm. The corresponding diffusion length due to thermally induced diffusion is:

$$L_d = \sqrt{2nD(T)t} \tag{4}$$

In the following we show that low energy ion-irradiation can be treated in a very similar way. Materials exposed to ion irradiation exhibit significant atomic rearrangement, so-called ion-beam mixing. Several processes are responsible for the ion mixing effect, all of which are initiated by the interaction of energetic ions with the solid. The relative significance of the ballistic, recoil and cascade effects can be altered by changing the mass and/or energy of the ions impinging on the sample. Increasing the mass of the ions increases the amount of energy deposited in nuclear collisions per unit length traveled by the ion. Consequently, the amount of mixing, $Q$, at the interface of two layers can be expressed as [21]:

$$Q \propto \left[\phi\left(\frac{dE}{dx}\right)_n\right]^{1/2} \tag{5}$$

Where $(dE/dx)_n$ is the nuclear stopping power and $\phi$ represents the ion dose that have passed through the interface. Electronic interaction between the beam and the atom results only ionization and causes no mixing hence it doesn’t play role in our case. Since the dose rate, in ion/cm$^2$/s, can be considered constant during an ion mixing experiment, the ion dose is proportional to time, leading to the observed behavior that mixing is proportional to (dose)$^{1/2}$ [22]. This implies that ion mixing is also proportional to the square root of ion mixing time. One can compare this proportionality to that observed for a reaction layer formed between two materials by thermally activated interatomic diffusion [23]. The width of the reacted layer *W*, found to behave as follows:

$$W \propto \sqrt{2\tilde{D}t} = L_d \tag{6}$$

where $\tilde{D}$ is of the form of an interdiffusion coefficient. Because of these similarities many ion mixing models are based on diffusion and interdiffusion analogues with mixing described in terms of an effective mixing diffusion coefficient, $\tilde{D}$.

## III EXPERIMENT

Highly ordered (majority $L1_0$) and also disordered (fcc) isotope-periodic $Fe_{47}Pd_{53}$ were grown on 20×20×2 $mm^3$ MgO(001) substrates by the method of molecular-beam epitaxy (MBE). In the case of ordered sample the substrate was held at 350°C and in the case of disordered sample the substrate was at room temperature during growth. In order to obtain epitaxial growth of Pd, a seed layer of 3 nm Cr was evaporated onto the MgO(001) surface at a rate of 0.07 Å/s, followed by the growth of a Pd buffer layer of 15 nm thickness at a rate of 0.2 Å/s. Then the bi-layer sequence of 3 nm $^{nat}Fe_{47}Pd_{53}$ and 2 nm $^{57}Fe_{47}Pd_{53}$ was repeated ten times. All $^{nat}Fe_{47}Pd_{53}$ and $^{57}Fe_{47}Pd_{53}$ layers were prepared by co-evaporation of Fe and Pd at a rate of 0.0485 Å/s for both $^{57}Fe$ and $^{nat}Fe$ and 0.0682 Å/s for Pd. To avoid oxidation of Fe a 1 nm Pd layer was grown on top of the sample. The $^{57}Fe$ was evaporated from Knudsen-cell (at 1410 $^{o}$C), all other layers were deposited using electron-beam evaporation. The base pressure was $2\times10^{-10}$ mbar in the MBE chamber, which raised to $2.8\times10^{-9}$ mbar during the growth. It is very important to achieve identical layer composition in the $^{nat}FePd$ and $^{57}FePd$ layers, therefore the so-called MBE tooling factors were carefully calibrated using Fe and Pd as well as $Fe_xPd_y$ test films, the layer thickness and the composition of which were determined by x-ray reflectometry and Rutherford Back-Scattering. The evaporation was controlled by two independent quartz thickness monitors.

The samples were cut into eight equal pieces of 10×5×2 $mm^3$. Before $He^+$-ion irradiation, SRIM [18] simulation were performed and the energy of the ion was chosen so that the $He^+$ passes through the layer and stops deep in the substrate without causing collision cascades. The ordered samples were then irradiated by $1.0\times10^{14}$ to $1.49\times10^{16}$ at/$cm^2$ and the disordered ones were irradiated by $3.7\times10^{14}$ to $3\times10^{16}$ at/$cm^2$ $He^+$ of energy 130 keV and one sample of each set were left unirradiated.

The CEMS experiments were performed by using a $^{57}$Co(Rh) single line Mössbauer source with a home made gas-flow single-wire proportional counter operating with He gas with 4.7% $CH_4$ extinction gas and at a bias voltage of 830±10 V.

SMR measurements were performed at the ID18 and ID22 beam line of the European Synchrotron Radiation Facility (ESRF, Grenoble, France) in 16 buntch mode. For sequential monochromatization of the beam to the 14.4 keV Mössbauer transition (($\lambda$=0.86025 Å) of $^{57}$Fe, a Si(111), then a Si(422)/Si(12.2.2) double channel cut monochromator were used. Prompt (non-resonant) and $^{57}$Fe delayed (time integrated) nuclear resonant reflectograms were recorded on each sample. In this method the interacting photon scatter on both the atomic electron and nucleus. The characteristic time of the electronic scattering is typically $10^{-15}$s from which the signal of nuclear scattering is well distinguishable since it is seven orders slower (~$10^{-8}$s). This means that the delayed signal is fully originates from the nuclear scattering and basically free from noise. Since the monochromatization of the synchrotron beam is currently a few meV and this is orders of magnitude higher than the transition energy of the nucleus (~100 neV) the different transitions of the nucleus are excited together and this results interference in the time spectrum. This time spectrum can be considered as the Fourier transformation of the conventional Mössbauer spectrum. The Synchrotron Mössbauer Reflectometry is the integration of the time spectrum at each angle which gives a very useful isotope sensitive method.

X-ray diffraction was carried out using a D8 Discover type diffractometer (equipped with a Göbel-mirror on the primary side) in Bragg-Brentano geometry using Cu $K_\alpha$ radiation ($\lambda$=1.5415 Å).

## IV RESULT AND DISCUSSION

The $He^+$ irradiation not only causes interface mixing in the FePd samples, but also induces, as we shall see, structural transformation. Detailed CEMS, XRD and TEM evaluation of the ordered irradiated samples has been published elsewhere [24,25]. Our earlier CEMS experiment show, three distinct micro structural units, the low hyperfine (hf) field, a large hf field, and an intermediate hf field local environment could be identified in the ordered samples, which were attributed to the ordered $L1_0$, one iron-rich phase and the disordered fcc structural units, respectively (Figure 1 a.). The iron-rich environment consists of iron nanoclusters within the $L1_0$ matrix, being magnetically coupled with it. In the ordered and iron-rich components the hf field points out of the sample plane, while the disordered phase has random magnetic orientation. By increasing the fluence of the $He^+$ irradiation from zero to $14.9\times10^{15}$ ions/cm$^2$, the CEMS spectral fraction of the ordered $L1_0$ phase decreases from 81% to 44%, while the disordered phase increases from 15% to 36%. The fraction of the partially ordered component also increases from 3.4% to 20%, but it reaches this value already at a fluence of $7.4\times10^{15}$ ions/cm$^2$ (Figure 2 a.)

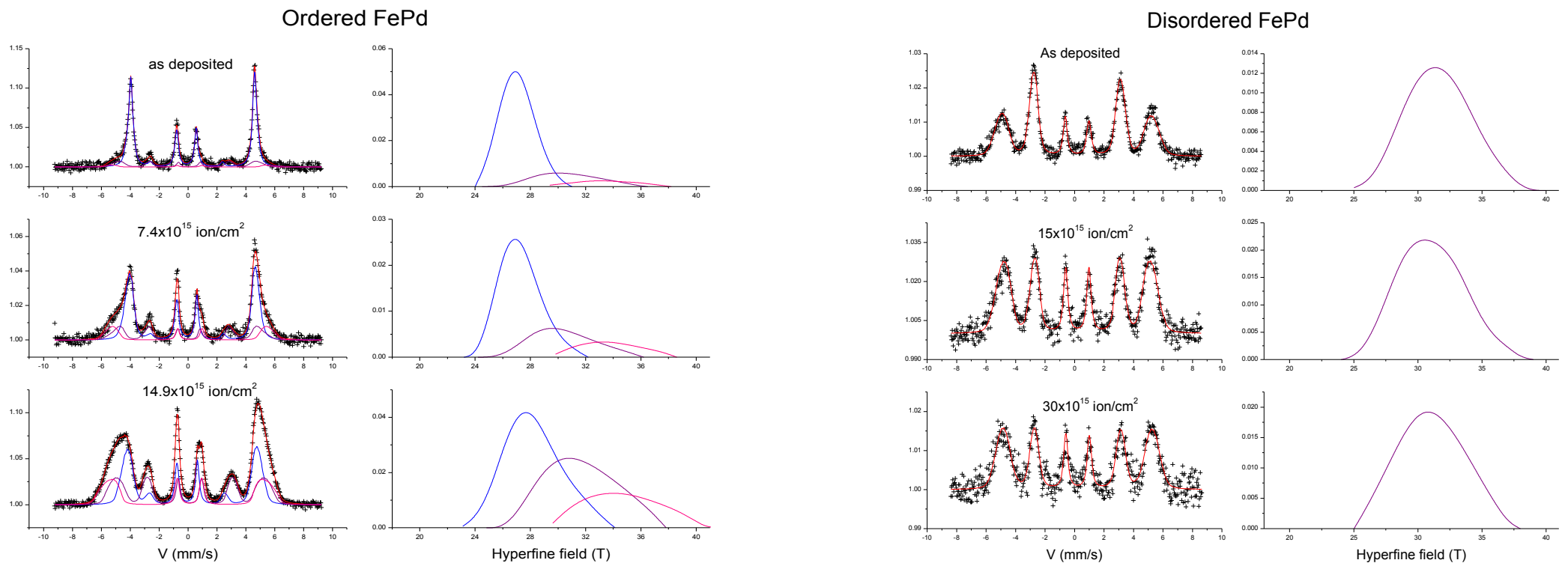

**Figure 1** The change of the CEMS spectra and hyperfine field distribution taken after growth, medium and high irradiation. A) ordered sample b) disordered sample.

The CEMS characterization of the disordered samples showed the lack of the $L1_0$ and the high hf field component (Figure 1 b.) and only the disordered fcc component is present in these kind of samples. The irradiation does not effent this structure (Figure 2 b.).

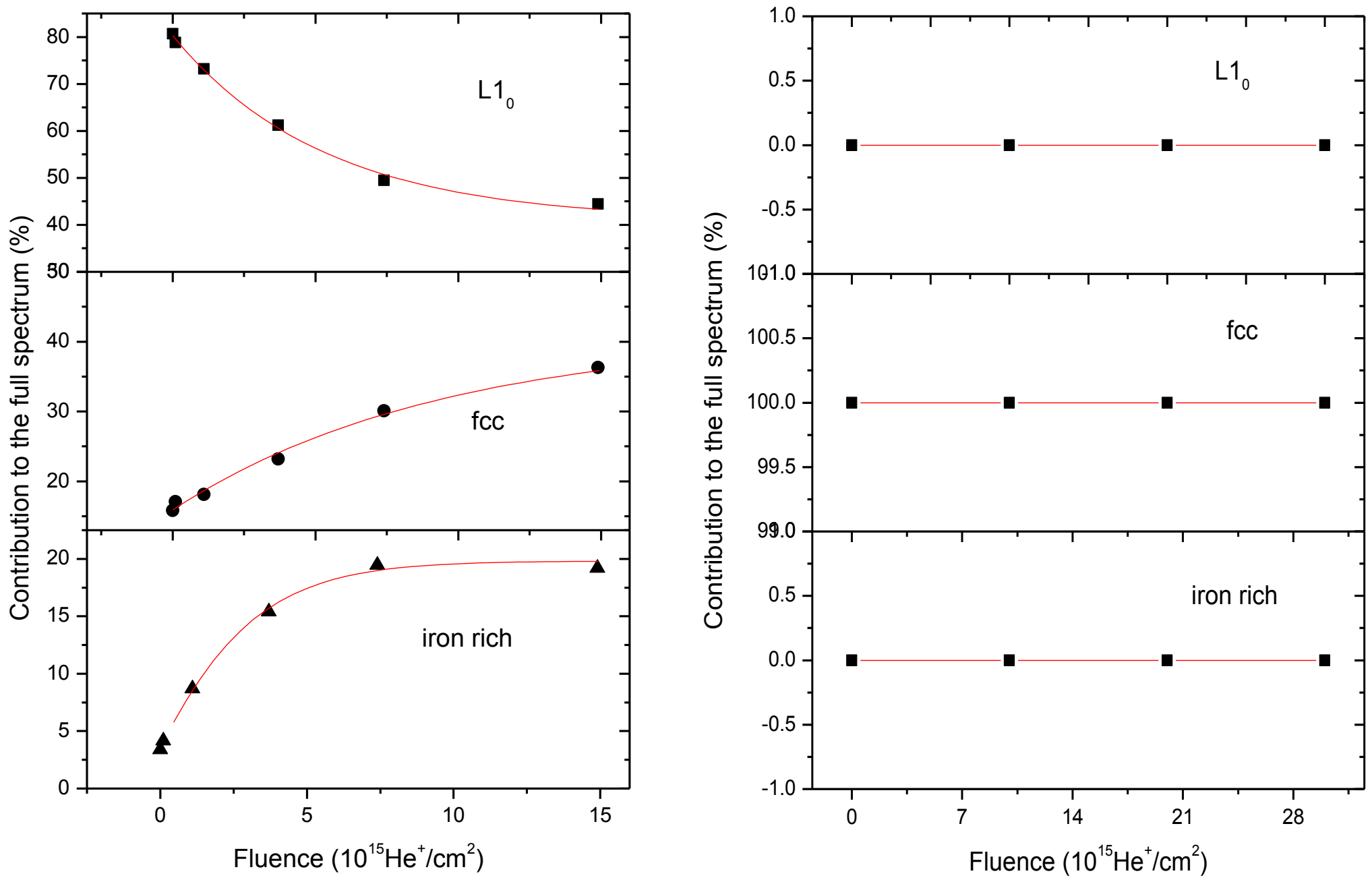

**Figure 2** The change of fractions of the different phases in ( a) ordered b) disordered) FePd after various fluences of $He^+$ irradiation as obtained from CEMS spectral intensities of the different species. a) The ordered $L1_0$ phase decreases from 81% to 44%, while the disordered phase increases from 15% to 36%. The fraction of the iron ri component increases from 3.4% to 20%. b) The fraction of the disordered fcc environment increased from 52% to 62% and the iron region decreased from 48% to 38%.

In order to gather information about the distribution of the local environment obtained from CEMS, additional selected area electron diffraction (SAED) measurements on the ordered sample were performed. This method is highly suitable to distinguish separate phases since it works in reciprocal space and so the higher index reflections of similar structures can be well separated On Figure 3 SAED measurement taken on the full sample (fcc-Pd + $L1_0$) and nano beam diffractin taken only on $L1_0$-FePd are compared.

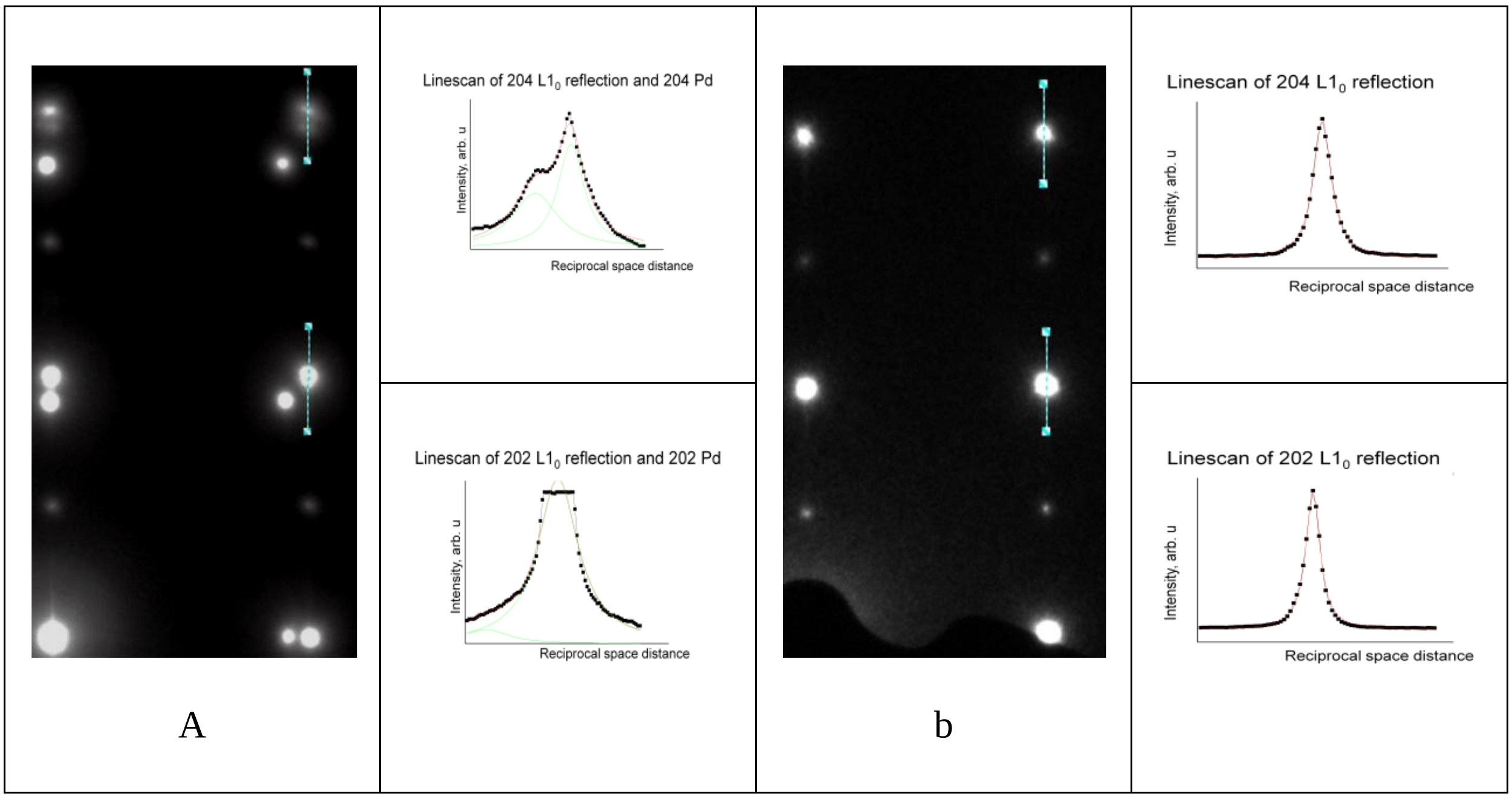

**Figure 3** Selected area electron diffraction taken from MgO, Cr, Pd and FePd layers (a) and the corresponding linescans taken on $L1_0$(204) and Pd(204) (up) and on $L1_0$(202) and Pd(202) (down). Nano-beam diffraction taken from only FePd layers (b) (from an area of ~10nm in diameter) and the linescan of $L1_0$(204) (up) and $L1_0$(202) down.

When distinct fcc-FePd phases separated by phase boundaries were present in the $L1_0$-FePd structure then it could be seen on the distribution of the 204 index reflection. of the SAED snapshot taken from the FePd layer, Since no such distribution is present we concluded that the size of any kind of precipitate cannot exceed the size of 5 Å and the structure of the sample can be treated as a random alloy of separate microstructural environments.

The evaluation of the reflectivity spectra in similar previous studies [[16,17 ,26]] was restricted to follow the decay of the isotopic multilayer (ML) Bragg peak height in the time integral SMR reflectivity curve normalized to the intensity of the total reflection peak. However, since the Bragg peak shape depends on the hyperfine fields (and their distributions) of the different species in the layers [see also 27], it is not justified to take only the normalized Bragg peak amplitude into account. Moreover, the normalization - in general – is only possible to the intensity of the total reflection peak, as an “internal standard”, but its intensity is extremely sensitive not only to small differences in the successive beam alignments prior to recording the SMR spectra of the different samples, but also, due to the limited penetration of the radiation, to the absorption of the near-surface layers, which may be differently effected by the progress of the diffusion. Consequently, normalization of the isotope Bragg peak height to the total reflection peak height may become rather uncertain. Therefore the in a proper evaluation the entire reflectivity curve has to be fitted to a layer model, which takes the diffusion prehistory of the multilayer properly into account. This latter procedure was followed here. The evaluation of the delayed SMR spectra was performed by FitSuite [28] program (Figure 4). The implemented routine is organized as follows.

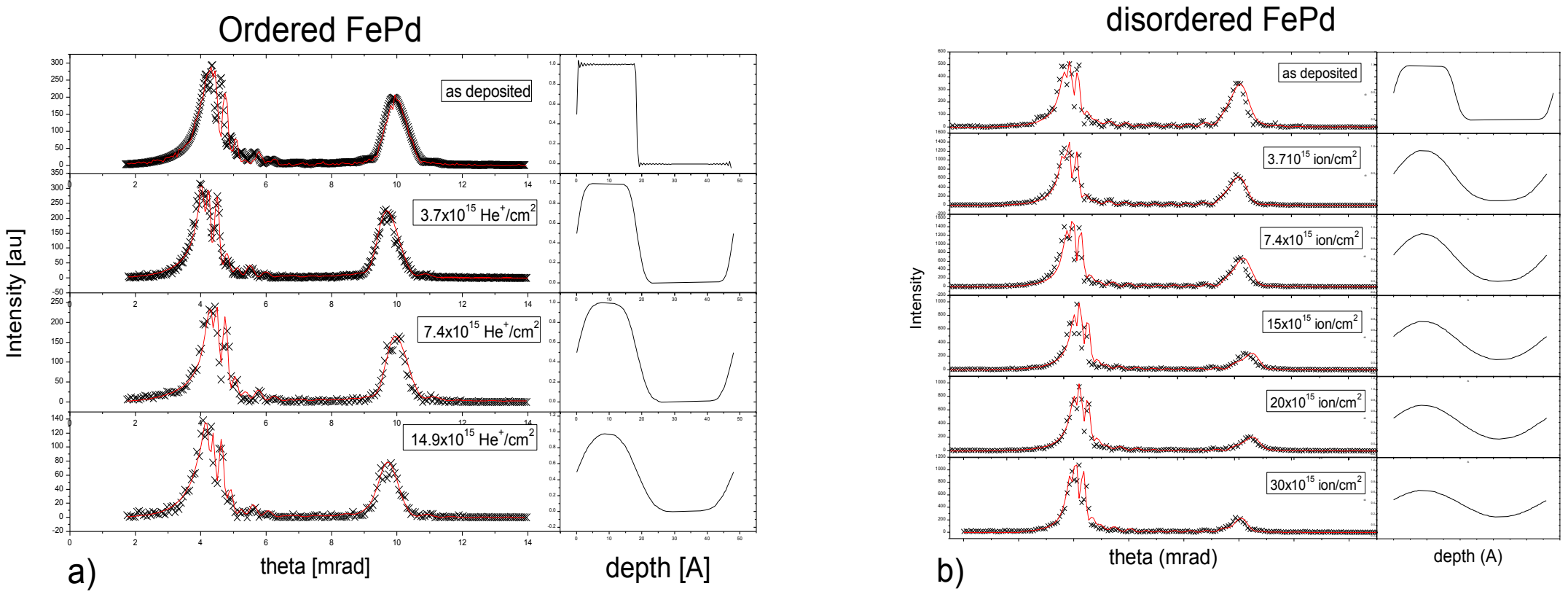

**Figure 4** The change of synchrotron Mössbauer refletivity spectra and the corresponding diffusion profile a) ordered FePd b) disordered FePd upon $He^+$ irradiation.

The squared diffusion lengths, $W_i^2$ can be summed up for successive heat treatments or consecutive irradiation steps. According to TEM studies the irradiated layers do not show distinct boundary-separated phases, therefore the Mössbauer species represent various environments in an otherwise single phase sample. Therefore, in the following we use the approximation of a multi-component random alloy. Figure 5 a and b shows the *Dt* evaluated from fitting SMR spectra for the FePd multilayers with different initial structure. On the left side the originally ordered sample while on the right side the originally disordered sample is present.

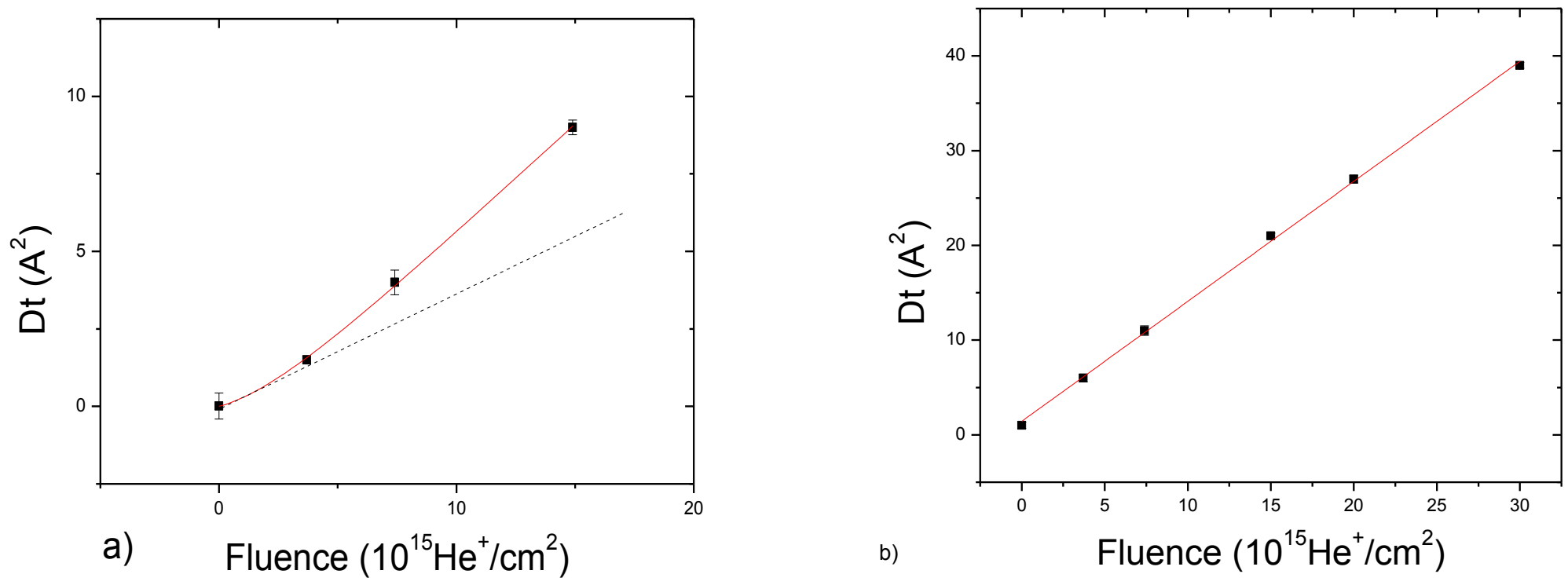

**Figure 5** The change of the fitted Dt for the whole system after $He^+$ irradiation. A) ordered b) disordered sample. The dashed line indicates the Dt for a structurally homogenous system.

At the as deposited samples there was only negligible mixing between the adjacent isotope layers. By the fulence of $15\times10^{15}$ $He^+/cm^2$ the Dt increased upto 9 Å$^2$ in the case of the ordered sample and reached 21 Å$^2$ for the disordered sample. This significant difference indicates that the inner structure of FePd plays major role in the scale diffusion process. One can also see that the two curves show different tendencies with the fluence of irradiation. It has been shown already that the effective mixing diffusion coefficient of a structurally homogenous material depends linearly on the irradiation dose [29]. By the ordered sample the growing tendency of the *Dt* exhibits a slight increase which is not the case of disordered FePd where it shows a constant tendency with the dose. The reason for the increasing Dt is that the ratio of the component which blocks the diffusion are decrasing while the ones in which the diffusion is faster increase after irradiation.

If we consider the model where the $\tilde{D}$ piecewise constant, thus $\tilde{D}$ takes an arbitrary positive value $\tilde{D}_i$ in volumes of arbitrary size, the squared diffusion length of Fe in the multilayer, $W \propto (\tilde{D}t)^{1/2}$ is, like in a random alloy, a weighed sums of the individual squared diffusion lengths of the different Fe-environments, $W_i = K_i(\tilde{D}_i t)^{1/2}$, i.e. the total *Dt* can be written as follows.

$$\tilde{D}t = x_{L1_0}\tilde{D}t_{L1_0} + x_{fcc}\tilde{D}t_{fcc} + x_{PO}\tilde{D}t_{PO} \tag{6}$$

Where $x_{L1_0}$, $x_{fcc}$, $x_{PO}$ and $Dt_{L1_0}$, $Dt_{fcc}$, $Dt_{PO}$ are the ratios and the effective diffusion parameters of the ordered $L1_0$, disordered fcc and partially ordered phases, respectively. The ratios as well as the effective diffusion parameters depend on the dose of irradiation. The fluence dependence of the ratios were determined by fitting (see Figure 2).

With these conditions (6) has the following form:

$$\tilde{D}t(\phi) = (\tilde{D}t^0_{L1_0} + a\phi)\frac{1}{\phi}\int_0^\phi x_{L1_0}(\phi)d\phi + (\tilde{D}t^0_{fcc} + b\phi)\frac{1}{\phi}\int_0^\phi x_{fcc}(\phi)d\phi + (\tilde{D}t^0_{PO} + c\phi)\frac{1}{\phi}\int_0^\phi x_{PO}(\phi)d\phi \quad (7)$$

By simultaneously fitting the total *Dt* of the ordered and disordered system obtained from evaluation of SMR spectra (Figure 4) with (7) we get the contribution of each particular environment to the diffusion of the entire multilayer (Figure 6).

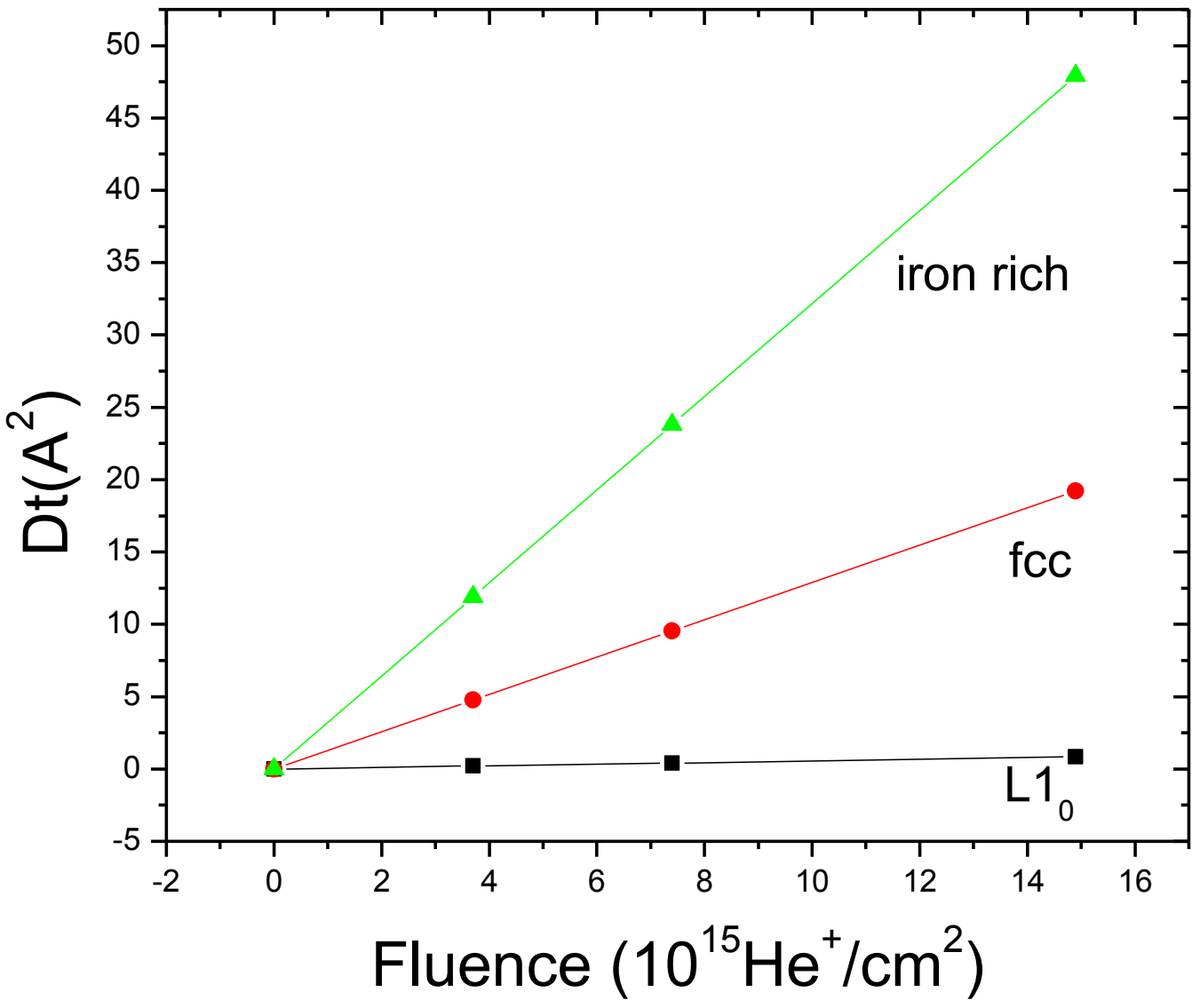

**Figure 6** Contribution of the individual Fe-environments in the FePd to the total diffusion length

The ratio of the effective diffusion parameters in the $L1_0$, fcc and partially ordered structures are 1:42:64, respectively. The effective diffusion lengths (perpendicular to the sample plane) in the different microstructural species are significantly different. Indeed, in the ordered $L1_0$ phase the diffusion of iron is almost blocked (the direction perpendicular to the sample plane coincides with the crystallographic *c* axis in these epitaxial samples). This means, that atomic migration across $^{57}$FePd/$^{nat}$FePd “interface” occurs mainly via the iron rich regions. According to (4), the diffusion lengths are related as 1:6.4:8 in the $L1_0$, the fcc and the iron rich regions, respectively. Taking a plausible assumption, namely that the value of the diffusion coefficient in the *fcc* phase is between the diffusion coefficient of the $L1_0$ phase in the crystallographic *c*-direction and the diffusion coefficient of the $L1_0$ phase in the *a-b* plane, the diffusion coefficient in the *c*-direction of the $L1_0$ phase is more than 1.9 times lower than in the *a-b* plane. Indeed, vacancy diffusion in the *c*-direction requires exchange of a vacancy between the Fe layer and the Pd layer, which have rather different vacancy creation energies.

## V. SUMMARY

In summary, the short range diffusion of $^{57}$Fe in highly ordered and disordered isotope-periodic epitaxial $Fe_{47}Pd_{53}$ thin film after varying fluence of 0 to $14.9\times10^{15}$ ion/cm$^2$ in the ordered and 0-30 ion/cm$^2$ of 130 keV $He^+$ irradiation in the disordered sample was investigated. The local structural changes were followed by Mössbauer spectroscopy. With increasing fluence, in the case of the ordered FePd, the fraction of the disordered and of the partially ordered components increased at the expense of the ordered $L1_0$ phase. In the case of disordered sample the $L1_0$ structure was not present and the fcc region increased at the expense of the iron rich region. The evaluation of the full delayed SMR intensity curves showed considerable interdiffusion of the isotope layers. The total diffusion length of the multiplayer is separated to individual diffusion lengths of the distinct microstructural species of the otherwise homogeneous phase. By fitting the variation of effective diffusion parameters using the corresponding Mössbauer spectral intensities of the individual Fe-environments as weights, we obtained an effective diffusion length for each Fe-environment. We find the ratio of the effective diffusion lengths in the $L1_0$, fcc and partially ordered structure to be 1:6.4:8, respectively By assuming the diffusion coefficient of the *fcc* phase to be between that of the $L1_0$ phase in the crystallographic *c*-direction and perpendicular to it, the diffusion coefficient in the *c*-direction of the $L1_0$ phase is more than 6.4 times lower than in the *x*-*y* plane.

## ACKNOWLEDGEMENTS

This work was partially supported by the Hungarian National Fund (OTKA) and by the National Office for Research and Technology of Hungary under contract numbers K 62272 and NAP-VENEUS'05, respectively. Beam time and ESRF